\documentclass[aps,twocolumn,showpacs,superscriptaddress,nofootinbib]{revtex4-2}

\usepackage{graphicx}
\usepackage{latexsym}

\usepackage{color}

\begin{document}
\title{
Preaveraging description of polymer nonequilibrium stretching
}

\author{Takuya Saito}
\email[Electric mail:]{tsaito@phys.aoyama.ac.jp}
\affiliation{Department of physical sciences, Aoyama Gakuin University, Chuo-ku, Sagamihara 252-5258, Japan}

\def\Vec#1{\mbox{\boldmath $#1$}}
\def\degC{\kern-.2em\r{}\kern-.3em C}

\def\SimIneA{\hspace{0.3em}\raisebox{0.4ex}{$<$}\hspace{-0.75em}\raisebox{-.7ex}{$\sim$}\hspace{0.3em}} 

\def\SimIneB{\hspace{0.3em}\raisebox{0.4ex}{$>$}\hspace{-0.75em}\raisebox{-.7ex}{$\sim$}\hspace{0.3em}}

\date{\today}

\begin{abstract}
This article focuses on a preaveraging description of polymer nonequilibrium stretching, where a single polymer undergoes a transient process from equilibrium to nonequilibrium steady state by pulling one chain end.
The preaveraging method combined with mode analysis reduces the original Langevin equation to a simplified form for both a stretched steady state and an equilibrium state, even in the presence of self-avoiding repulsive interactions spanning a long range.
However, the transient stretching process exhibits evolution of a hierarchal regime structure, which means a qualitative temporal change in probabilistic distributions assumed in preaveraging.
We investigate the preaveraging method for evolution of the regime structure with consideration of the nonequilibrium work relations and deviations from the fluctuation-dissipation relation.
\end{abstract}

\pacs{05.40.-a,05.10.Gg,82.35.Lr,83.80.Rs}

\def\degC{\kern-.2em\r{}\kern-.3em C}

\newcommand{\gsim}{\hspace{0.3em}\raisebox{0.5ex}{$>$}\hspace{-0.75em}\raisebox{-.7ex}{$\sim$}\hspace{0.3em}} 
\newcommand{\lsim}{\hspace{0.3em}\raisebox{0.5ex}{$<$}\hspace{-0.75em}\raisebox{-.7ex}{$\sim$}\hspace{0.3em}} 

\def\Vec#1{\mbox{\boldmath $#1$}}

\maketitle

\section{Introduction}

Ubiquitous phenomena from small to large scale observed in the materials and life sciences through to the geosciences are discovered as stochastic systems~\cite{StatisticalPhysics_II,PhysToday_Barkai_2012,SoftMatter_Sokolv_2012,vanKampen,Gardiner,SekimotoBook,GeophysResLett_Ide_2008,JPhys_Schiessel_2003,Mandelbrot}, of which appropriate simple descriptions may be formulated to meet with spatiotemporal resolutions.
In dealing with systems that have large internal degrees of freedom, 
discerning noise in a coarse-grained picture and also in the structural regimes based on scale separation is crucial.

Soft matter, as exemplified by a polymer, may exhibit viscous or elastic motion depending on the spatiotemporal scale, and involves some characteristic scale quantities that classify multiple regimes.
Let us consider a single polymer.
A dynamical regime is intimately associated with static spatial structures. 
A global structure forms a fractal referred to as a random coil that does not rely on local specificities such as proper chemical bonds~\cite{deGennesBook,Doi_Edwards,Doi_SoftMatter}.
The fractal regime creates a remarkable dynamical feature called anomalous diffusion, which is defined as the nonlinear growth of the mean-square displacement (MSD)~\cite{PhysToday_Barkai_2012,SoftMatter_Sokolv_2012}, while a Brownian particle exhibits only normal diffusion that is linearly proportional to time.
An important issue in polymer physics is to incorporate long-range repulsive interaction that yields nonlinear self-avoiding (SA) effects, and mode analyses have been developed to allow the reproduction of polymer anomalous diffusion by employing a preaveraging method~\cite{Doi_SoftMatter,JChemPhys_Schiessel_Oshanin_Blumen_1995,JStatMech_Panja_2010,PRE_Lizana_Barkai_Lomholt_2010,PRE_Sakaue_2013,PRE_Saito_2015}, 
where the probability distribution in variables other than those of interest is assumed in advance~\cite{Doi_Edwards}.

In the equilibrium for a simple polymer, the global structure has a few stationary fractal regimes.
However, the nonequilibrium conditions do not ensure the same case, as illustrated in polymer stretching (see fig.~1)~\cite{EPL_Brochard_1993,Macromolecules_Marciano_Brochard_1995}.
Turning our attention to SA polymers with nonlocal interaction,
we are aware that polymer stretching undergoes a qualitative temporal change in the fractal regime structure due to the effective emergence of a distinct interaction range~\cite{PRE_Sakaue_2007,PRE_Sakaue_2012,PRE_Saito_2015,PRE_Saito_2012,PRE_Saito_2017}, which should modify the distribution assumed in preaveraging.
This article focuses on the preaveraging description relevant to the temporal evolution of the fractal regime structure~\cite{EPL_Brochard_1993,Macromolecules_Marciano_Brochard_1995}.
This is an interesting issue in stochastic energetics~\cite{SekimotoBook} because an interpretation of the energy balance is inherent in the resolution scales, while multiple scales involve the polymer stochastic processes.

The manuscript is organized as follows.
Section~\ref{stretching} introduces polymer stretching with a concept for the evolution of the regime structure, and then we give the mode description with an effective Langevin equation based on preaveraging.
We then discuss the nonequilibrium work relation~\cite{PRL_Jarzynski_1997,CRPhysique_Jarzynski_2007,JETP_Bochkov_Kuzovlev_1977,JETP_Bochkov_Kuzovlev_1979,PRE_Hatano_1999,PRL_Hatano_Sasa_2001} and the deviations from the fluctuation-dissipation relation (FDR)~\cite{PRL_Harada_Sasa_2005,PRE_Harada_Sasa_2006,EPL_Speck_Seifert_2006,PRE_Deutsch_Narayan_2006}, which play a crucial role in the progress on nonequilibrium physics at the small scale.
Section~\ref{discussion} discusses applications and perspectives.
Finally, we summarize the study in Section~\ref{conclusion}.

\begin{figure}[ht]
\begin{center}
\includegraphics[scale=0.45]{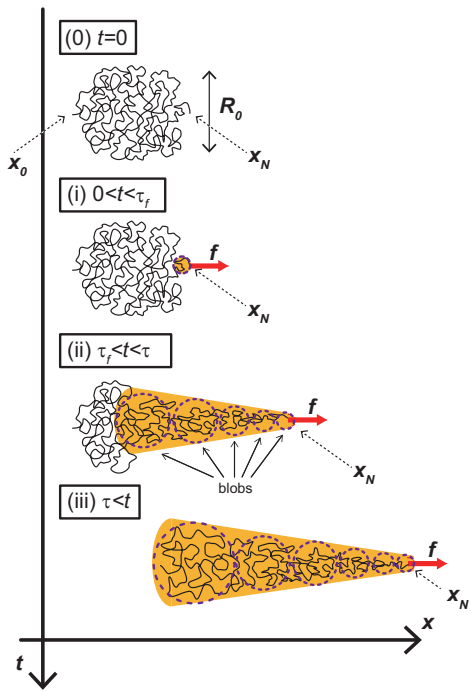}
      \caption{
		(Color online) Schematic representation of polymer stretching.
		(0) initial equilibrium coil. 
		(i) early blob formation for $0<t<\tau_f$.
		(ii) stretching transient for $\tau_f<t<\tau$.
		(iii) nonequilibrium steady state for $t>\tau$.
		Dashed circles represent blobs (local coilings), and an entire moving domain is shaded.
	  }
\label{fig1}
\end{center}
\end{figure}

\section{Evolution of regime structure}
\label{stretching}

We begin with the physical mechanism of polymer stretching~\cite{PRE_Sakaue_2007,PRE_Sakaue_2012,PRE_Saito_2015,PRE_Saito_2012}, as shown by the schematic representation in fig.~1.
A linear homopolymer chain is suspended in a solution that serves as a thermal bath.
(0) The polymer is initially in equilibrium.
One chain end starts being pulled with force magnitude $f$ at $t=0$.
(i) At the beginning, only the monomers close to the forced monomer along the chain are set in motion, and then the moving domain (shaded in fig.~1\,(i)), where tension has propagated, grows in time while qualitatively retaining the equilibrium shape.
However, a force magnitude larger than a threshold $f\simeq k_BT/R_0$ is applied, so that the polymer cannot sustain the equilibrium shape for approximately $t>\tau_f$.\footnote{Relaxation time for a monomer unit or for a blob is denoted by $\tau_u$ or $\tau_f \simeq \tau_u (\xi_f/a)^z$, respectively. 
Note that $z$ is referred to as the dynamical exponent that relates the relaxation time to the correlation length $\xi_f$ (see also the paragraph after eq.~(\ref{coeff_q0})).}
Note that $R_0$ denotes the spatial size of the equilibrium polymer (see fig.~1\,(0)).
(ii) The polymer is largely deformed, i.e., the nonequilibrium stretched shape appears around $t=\tau_f$. 
The moving domain (shaded in fig.~1\,(ii)) grows, accompanied by the propagation of tension along the chain, i.e., undergoing a transient stretching process. 
(iii) When the tension eventually arrives at the other chain end, the entire polymer enters a stretching nonequilibrium steady state~\cite{EPL_Brochard_1993,Macromolecules_Marciano_Brochard_1995}.

There are a few scale-dependent regimes to express the static conformation for equilibrium (see (0) and (i)): (a) local specific structures smaller than the monomer (or less than the Kuhn length);
 (b) global fractals, for which asymptotic behavior does not rely on local specific characteristics, such as the chemical bonds.
Note that this article considers only a flexible polymer for simplicity; therefore, we may treat just a single fractal regime in the equilibrium.\footnote{For a semiflexible polymer with local bending stiffness such as double-stranded DNA, two fractal regimes may be observed due to the SA effects, even in equilibrium dependent on the chain length~\cite{deGennesBook}. Using the Flory exponent $\nu$, the short polymer chain assumes $\nu=1/2$, while the longer polymer chain assumes those of the SA polymer, i.e., $\nu=3/4$ or $\nu\simeq 0.588...$ in two or three dimensions, respectively.}
On the other hand, the nonequilibrium stretching state exhibits a different hierarchal organization of a regime structure, which adds one more level (see (ii) and (iii)).
The fractal structure, in a broad sense, is represented by a blob (dashed circle) at the intermediate scale and global one-dimensional deformation is represented by a sequence of these blobs (the entire shaded domain). 
Therefore, the stretching transient process intrinsically reorganizes the regime structure.

\subsection{Formulation}
\label{formalism}

We formulate the large-scale dynamics with the Langevin equation.
The linear polymer is written with $N+1$ monomer units being $a$ in each size.\footnote{
The way of counting determines the total number of monomers; however, a difference between $N$ and $N+1$ does not cause a significant problem in the long chain limit of interest.}
Monomer indices are denoted by $n$ assigned from a chain end.
A subscript $n$ indicates a variable of the $n$-th monomer; e.g., $x_n(t)$ is the position of the $n$-th monomer. No subscript indicates a variable of the forced monomer $x(t)\equiv x_N(t)$ (the $N$-th monomer is pulled).
In addition, for compact notation, we introduce a difference from the time origin or the mean value, which are respectively denoted by
\begin{eqnarray}
\Delta x(t)\equiv x(t)-x(0), \quad \delta x(t) \equiv x(t)-\left<x(t)\right>,
\label{def_diff}
\end{eqnarray}
where $\left< (\cdot) \right>$ denotes the ensemble average.

An elementary expression for overdamped Langevin motion in real space is translated into the preaveraged description in mode space:
\begin{eqnarray}
\sum_{n'} \gamma [x_{n},x_{n'}] \frac{dx_{n'}(t)}{dt}
=
-\frac{\partial {\cal E}_{(e)}(\{ x_{i} \})}{\partial x_{n}}
 +f_n(t)+\zeta_n(t).
\label{real_original_EM}
\end{eqnarray}
\qquad \qquad \qquad \qquad \qquad \qquad $\Downarrow$
\begin{eqnarray}
\gamma_q(t) \frac{dX_q(t)}{dt}
=
-\frac{\partial {\cal E}_q(X_q,k_q(t))}{\partial X_q}\Biggr|_{X_q(t)} +F_q(t)+Z_q(t).
\label{imag_EM_TDSA}
\end{eqnarray}
A transformation rule between the real space variable $x_n(t)$ and the normal mode $X_q(t)$ is defined as in ref.~\cite{Doi_Edwards}:
\begin{eqnarray}
X_q(t) \equiv \int_{0}^Ndn\, x_n(t) h_{q,n}, \quad x_n(t)=\sum_{q=0}^N X_q(t)h_{q,n}^\dagger,
\label{variable_normal}
\\
h_{q,n}\equiv \frac{1}{N}\cos{\left( \frac{qn \pi}{N} \right)},\quad h_{q,n}^\dagger = c_q^{-1} \cos{\left( \frac{\pi nq}{N}\right)},
\label{kernel_h}
\end{eqnarray}
after taking the continuum limit for $n$
(refer to appendix A or refs.~\cite{PRE_Saito_Sakaue_2017,PRE_Saito_2017,Polymers_Saito_2019}.\footnote{
Technically, the conversion rule in refs.~\cite{PRE_Saito_Sakaue_2017,PRE_Saito_2017,Polymers_Saito_2019} (by factors) is defined in a different way from the conventional rule that appears in the textbook of ref.~\cite{Doi_Edwards}.})

The left-hand side of eq.~(\ref{real_original_EM}), $\gamma [x_{n},x_{n'}]$ denotes the original (incompressive-fluid) frictional kernel with $x_n$ and $x_{n'}$, which may reflect the fluid-flow-mediated interactions.
There are two important models: free-draining, and non-draining.
The free-draining model employs local friction such as $\gamma [x_{n},x_{n'}]=\gamma \delta_{nn'}$ in a discrete expression, which means that Stokes drag is exerted on each monomer.
On the other hand, the non-draining model has a long-range hydrodynamic interaction, which yields friction heat due to the motion of distant monomers.
Regardless of the model,
in a random fractal polymer, the asymptotics of a two-point long-range correlation are replaced with a function of a contour distance along the chain: $\left<|x_n-x_{n'}|^2\right>\sim |n-n'|^{2\nu}$ with the Flory exponent $\nu$ (see after eq.~(\ref{coeff_q0}) for details).
The integral kernels are thus averaged in advance, or preaveraged.
This enables the use of an approximation such as $\gamma[x_{n},x_{n'}]\rightarrow \gamma_{n-n'}(t)$, which technically makes a convolution on the integral transform available, as reduced to eq.~(\ref{imag_EM_TDSA}).

On the right-hand side, the external driving force is $f_n(t)=f\delta_{nN}\Theta (t)$ with the Heaviside step function,
 and $F_q(t)$ is converted from $f_n(t)$ through eq.~(\ref{variable_normal}).
$\zeta_n(t)$ or $Z_q(t)$ denotes the Gaussian-distributed random noise with zero mean in the real or mode space, respectively, and the covariance is assumed to satisfy the FDR of the second kind $\left< Z_q(t)Z_{q'}(s) \right>=(2c_q\gamma_q(t)/N)k_BT\delta_{qq'}\delta (t-s)$ with correlation of the temporal delta-function in the mode space.
Note that the numerical factors $c_0=1$ for $q=0$ and $c_q=1/2$ for $q\geq 1$ are introduced.

The first term ${\cal E}_{(e)}(\{ x_{n} \})$ or ${\cal E}_q(X_q;k_q(t))$ on the right-hand side is called the effective Hamiltonian, which produces a conservative force between monomers as a function of $\{x_n\}$, where the subscript $(e)$ represents the elementary description before preaveraging.
A general polymer model necessarily has the bonding potential that connects monomers to form the chain. 
In this article, we also consider a repulsive potential between the monomers that gives rise to the SA effects, unless otherwise specified. 
A point in the approximation for the mode analyses is that, even under the long-range SA (and under hydrodynamic interaction), we may diagonalize the modal motion with harmonic potentials:
\begin{eqnarray}
{\cal E}_q(X_q;k_q(t)) = \frac{1}{2}k_q(t)X_q^2.
\label{SP_Uq}
\end{eqnarray}
Notation ${\cal E}_q$ with subscript $q$ is employed as the effective Hamiltonian for the $q$-mode component, which originally arises from the approximation 
${\cal E}_{(e)}(\{ x_{i} \}) \simeq \sum_q Nc_q{\cal E}_{q}(X_{q})(h_{q,N}^\dagger)^2$.
The mode diagonalization for the equilibrium SA polymer has been numerically verified as a good approximation~\cite{JCP_Panja_2009}.

The friction coefficient $\gamma_q(t)$ or the spring constant $k_q(t)$ may vary with time such that eq.~(\ref{imag_EM_TDSA}) can contain nonstationary dynamics.
Polymer stretching is well reproduced by specifying the time-dependent coefficients~\cite{PRE_Saito_2015} 
for the internal modes $q\geq 1$ with
\begin{eqnarray}
k_q(t)=
\left\{
 \begin{array}{ll}
  k (q/N)^{2\nu+1} & \mathrm{for}\ t<\tau_f
  \\
  \\
\left\{
 \begin{array}{ll}
 k^{(f)}(q/N)^2 & (q<q_f)
 \\
 k (q/N)^{2\nu+1} & (q>q_f)
 \end{array}
 \right. & \mathrm{for}\ t>\tau_f
  \end{array}
 \right.
 \label{spring}
\end{eqnarray}
\begin{eqnarray}
\gamma_q(t)=
\left\{
 \begin{array}{ll}
  \gamma (q/N)^{-\nu(z-2)+1} & \mathrm{for}\ t<\tau_f
  \\
  \\
\left\{
 \begin{array}{ll}
 \gamma^{(f)} & (q<q_f)
 \\
 \gamma(q/N)^{-\nu(z-2)+1} & (q>q_f)
 \end{array}
 \right. & \mathrm{for}\ t>\tau_f
  \end{array}
 \right.,
 \label{friction}
\end{eqnarray}
or for the center-of-mass mode $q=0$ with
\begin{eqnarray}
k_0(t)=0, \quad \gamma_0(t)=\gamma_1(t),
\label{coeff_q0}
\end{eqnarray}
where $k^{(f)}=k g_f/(\xi_f/a)^2\simeq k(fa/k_BT)^{(2\nu-1)/\nu}$, $\gamma^{(f)}\simeq \gamma (fa/k_BT)^{2-z+(1/\nu)}$ are the force-dependent components of the coefficients.
Here $\nu$ denotes the Flory exponent, which is used to express the polymer extension $R_0\sim N^\nu$ being the static properties,
while its inverse, i.e., $1/\nu$, is the fractal dimension in $N\sim R_0^{1/\nu}$.
$\nu=1/2$ for an ideal polymer chain, or $\nu=3/4$ and $\nu \simeq 0.588$ for the SA chain in the respective 2 and 3 dimensions~\cite{deGennesBook,Doi_Edwards,JCP_Panja_2009}.
On the dynamics side, the scaling of the characteristic relaxation time $\tau$ is expressed with the dynamical exponent $z$ as $\tau \sim R_0^z$ ($z=3$ for non-draining or $z=2+1/\nu$ for free-draining)~\cite{deGennesBook,Doi_Edwards}.
The top lines of eqs.~(\ref{spring}) and (\ref{friction}) for $t<\tau_f$ are the same of those for the equilibrium mode, as in ref.~\cite{Doi_Edwards,JCP_Panja_2009}.
The coefficients for $t>\tau_f$ capture the nonequilibrium characteristics relevant to the transient (ii) or the steady state (iii) regimes.
Note that $q_f=N(fa/k_BT)^{1/\nu}$ is a threshold mode index that separates the global nonequilibrium and intermediate equilibrium modes.\footnote{Let $\xi_f$ or $g_f$ be the blob size or the number of monomers in the blob, respectively.
The threshold mode number is estimated by $q_f\simeq N/g_f$ and $\xi_f \simeq a g_f^{\nu}$.}
Bear in mind that the characteristic time $t=\tau_f$ signifies the change in the distributions assumed in the preaveraging.

\subsection{Nonequilibrium work relations}
\label{terminal_diff}

An issue of nonequilibrium work relations is addressed here.
Small particles such as colloids suspended in a solution undergo stochastic motion induced by thermal agitation, where work, heat, and internal energy may be defined as the stochastic quantities according to the stochastic motion and the respective trajectories~\cite{SekimotoBook}.
For such thermal stochastic motion, 
the nonequilibrium work relations~\cite{PRL_Jarzynski_1997,CRPhysique_Jarzynski_2007,JETP_Bochkov_Kuzovlev_1977,JETP_Bochkov_Kuzovlev_1979} claim that averaging an exponential of work divided by thermal energy over the nonequilibrium processes between equilibrium states satisfies the integral relations, which is known as the Jarzynski equality~\cite{PRL_Jarzynski_1997}, or the Bochkov-Kuzovlev relation~\cite{JETP_Bochkov_Kuzovlev_1977,JETP_Bochkov_Kuzovlev_1979}, according to the work definition~\cite{CRPhysique_Jarzynski_2007}.
The former nonequilibrium average is identified with a free energy difference specified by the system equilibrium, and the latter relation takes the form of an identity.

In discussing application of the nonequilibrium work relation to the present system, there are some points to be aware of. 
The first point is house-keeping heat.
The stretching polymer finally arrives at the nonequilibrium steady state shape, in which the internal modal motions ($q\geq 1$) do not produce house-keeping heat.
On the other hand, the center of mass mode motion ($q=0$) keeps producing dissipative heat.
The internal modes could evolve as if they experienced the transition between the equilibrium states.
Accordingly, we are restricted to considering an application of the Jarzynski equality or the Bochkov-Kuzovlev relation to the internal modes,
while the Hatano-Sasa equality is the integral relation between (genuine) nonequilibrium steady states~\cite{PRE_Hatano_1999,PRL_Hatano_Sasa_2001}.

The second point concerns the definition of the work~\cite{PNAS_Hummer_Szabo_2001,PTP_Jarzynski_2006,CRPhysique_Jarzynski_2007}.
While the Bochkov-Kuzovlev relation employs mechanical work, as discussed later, our attention is first focused on an analogue of the Jarzynski equality by introducing the work defined as the change in the effective Hamiltonian varied by the external parameters:
\begin{eqnarray}
W(t)
&\equiv&
\int_0^tdt'\,
\sum_{q} 
c_qN
\Biggl[ 
\frac{\partial {\cal E}_q^{(f)}}{\partial k_q} \frac{dk_q(t')}{dt'} 
\nonumber \\
&&
\qquad 
+
\frac{\partial {\cal E}_q^{(f)}}{\partial F_q} \frac{dF_q(t')}{dt'} 
\Biggr](h_{q,N}^\dagger)^2,
\label{fic_work}
\end{eqnarray}
where the effective Hamiltonian is replaced so as to include the coupling term of the position with the external force:
\begin{eqnarray}
{\cal E}^{(f)}(\{ X_q;k_q,F_q\})
&\equiv&
{\cal E} -f(t)x(t)
\label{SP_Hqf}
\nonumber \\
&=&
\sum_{q\geq 1} c_qN{\cal E}_q^{(f)}(X_q,k_q)(h_{q,N}^\dagger)^2
\label{SP_Hq} 
\end{eqnarray}
with the $q$-mode components denoted by
\begin{eqnarray}
{\cal E}_q^{(f)}(X_q;k_q,F_q)&\equiv&{\cal E}_q-F_qX_q.
\label{SP_Hq_each_mode} 
\end{eqnarray}

The third point is the time dependence of the spring constants $k_q(t)$ for the SA polymer, while the nonequilibrium work relation is rigorously discussed for the Rouse polymer with a time-independent spring constant~\cite{EPJB_Speck_Seifert_2005,PRE_Dhar_2005,PRE_Sharma_Cherayil_2011}.
The second term in the brackets in eq.~(\ref{fic_work}) is the explicit work done by the applied force, which is literally given by changes in the external parameters.
On the other hand, we note that the work (eq.~(\ref{fic_work})) has non-conventional contributions. 
Considering the physical meaning of the nonequilibrium polymer stretching, the temporal variation in the spring constants should be envisaged as the spontaneous change, not the externally controlled change.
Nonetheless, we virtually consider the spring constants $k_q(t)$ as externally controlled parameters embedded in the effective Hamiltonian.

Using eqs.~(\ref{fic_work}) and (\ref{SP_Hqf}), the energy conservation for each stochastic process is written as
\begin{eqnarray}
\Delta {\cal E}^{(f)}(X_q(t);k_q(t),F_q(t)) &=& W(t)-Q(t),
\label{energy_conservation_law}
\end{eqnarray}
where heat is defined through the mode space as
\begin{eqnarray}
Q(t) &=& \sum_{q\geq 0} c_qN\left( \gamma_q \frac{dX_q}{dt}-Z_q(t)  \right) \circ \frac{dX_q}{dt}(h_{q,N}^\dagger)^2,
\label{SP_Qq}
\end{eqnarray}
with $\circ$ denoting Stratonovich multiplication.
Equation~(\ref{SP_Qq}) is motivated by the definition of the real space with Gaussian white noise~\cite{SekimotoBook}.
From the equation of motion (eq.~(\ref{imag_EM_TDSA})), we can confirm that the energy balance eq.~(\ref{energy_conservation_law}) holds under eqs.~(\ref{fic_work})--(\ref{SP_Hq_each_mode}) and eq.~(\ref{SP_Qq}).
Note that the heat flow from the system to the external environment is assigned to be positive in eq.~(\ref{SP_Qq}).

As with the first point, extraction of the internal mode components ($q\geq 1$) from $W$ and rewriting them as $W_{q\geq 1}$
leads to the expectation of an analogue to the Jarzynski equality~\cite{PRL_Jarzynski_1997,CRPhysique_Jarzynski_2007} for $t\gg \tau$:
\begin{eqnarray}
\left< e^{-W_{q\geq 1}/(k_BT)} \right>= e^{-\Delta \Phi/(k_BT)}.
\label{fic_W_F}
\end{eqnarray}
The ensemble average is taken over the initial configurations, and all the paths from the initial to the final equilibrium (or nonequilibrium steady) states.
In addition, $\Phi$ is referred to as the thermodynamic potential for the fixed-tension ensemble~\cite{StatisticalPhysics_II}, defined as follows:
\begin{eqnarray}
\Phi 
&=& 
-\sum_{q\geq 1} 
[c_qN(h_{q,N}^\dagger)^2] 
\frac{c_qk_BT}{N} 
\nonumber \\
&& \times 
\log{} \left[ \int_{-\infty}^{+\infty} dX_q\,\exp{\left(-\frac{{\cal E}_q^{(f)}}{c_qk_BT/N}\right)} \right],
\label{tension_free_energy}
\end{eqnarray}
where the factor $c_qk_BT/N$ corresponds to the effective temperature of the $q$-mode space (see the consistency with $\left< Z_q(t)Z_{q'}(s) \right>=(2c_q\gamma_q(t)/N)k_BT\delta_{qq'}\delta (t-s)$).
$\Delta \Phi=\Phi (t)-\Phi (0)$ represents a difference written as follows:
\begin{eqnarray}
\Delta \Phi &=&  
\sum_{q\geq 1}\Biggl[
-\frac{f^2}{4Nk_q^{(f)}} 
-\frac{k_BT}{4} \log{\sqrt{\frac{k_q}{k_q^{(f)}}}} \Biggr] (h_{q,N}^\dagger)^2,
\label{F_Wima}
\end{eqnarray}
where the first and second terms in brackets, which are derived from changes in the external force and in the spring constants, respectively, appear as additive forms (see appendix~B).
The present system evolves under Gaussian processes, of which the calculi for the ensemble average may be performed through path integrals with reference to ref.~\cite{PRE_Minh_Adib_2009,PathIntegrals_Chaichain_Demichev,Feynman_Hibbs}.

We next consider the distinct nonequilibrium work relation referred to as the Bochkov-Kuzovlev relation~\cite{JETP_Bochkov_Kuzovlev_1977,JETP_Bochkov_Kuzovlev_1979}, where the work is defined as mechanical work (i.e., force times displacement as conventionally found in textbooks of classical mechanics).
In the same way as in the integral relation of eq.~(\ref{fic_W_F}), the change in the spring constant must be incorporated into the mechanical work as a fictive part, and then the Bochkov-Kuzovlev relation written for the internal modes:
\begin{eqnarray}
\left< e^{-W_{q\geq 1}^{(0)}/(k_BT)} \right>= 1,
\label{mechanical_noneqW}
\end{eqnarray}
where the mechanical work for $q\geq 1$ is introduced as
\begin{eqnarray}
W_{q\geq 1}^{(0)}&=&
\int_0^{t}dt'\,\sum_{q\geq 1} 
c_qN
\Biggl[ 
-\frac{1}{2}\Delta k_q(t')
\frac{d[X_q(t')^2]}{dt'} 
\nonumber \\
&&
+
F_q(t')\frac{dX_q(t')}{dt'} 
\Biggr](h_{q,N}^\dagger)^2
\label{W_mechanical}
\end{eqnarray}
such that
\begin{eqnarray}
-W_{q\geq 1}
+
W_{q\geq 1}^{(0)}
&=&
\sum_{q\geq 1} c_qN\Biggl[ 
-\frac{1}{2} \Delta k_q(t) X_q(t)^2
\nonumber \\
&&
+
F_q(t) X_q(t)
\Biggr](h_{q,N}^\dagger)^2
\end{eqnarray}
with $dW_{q\geq 1}^{(0)}/dt|_{t=0}=0$ as in ref.~\cite{CRPhysique_Jarzynski_2007}.
Note that in the work definition of eq.~(\ref{W_mechanical}), $F_q(t)=F_q\Theta (t-\epsilon_+)$ with infinitesimal positive $\epsilon_+$, so that we can have $F_q(0)= 0$ required for an applicable condition in the Bochkov-Kuzovlev relation.

A short summary of this section is that 
the preaveraging description may become associated with the nonequilibrium work relations (eqs.~(\ref{fic_W_F}) and (\ref{mechanical_noneqW})) 
if we assume that the changes in the spring constants that do not intrinsically belong to the external parameters become the work (see appendices C and D for an observation from the elementary to the preaveraged Langevin equations).
In addition, the energy balance at each respective mode in the present regime structure, or at each respective hierarchal level with the characteristic scale, is independently interpreted
so that the interactions between the distinct modes or regimes may be reduced to the external parameter changes.

\subsection{Transient deviations from FDR}
\label{FDR_dev}

The next issue is the transient side on the regime structure relevant to the FDR.
Two integral forms of the FDR are utilized here:
(a) The first integral form is the equality often involved in anomalous diffusion by monitoring of the displacement $\Delta x(t)$.
The conventional approach is to observe the MSD $\left< \Delta x(t)^2 \right>$ as a primary statistic~\cite{JChemPhys_Schiessel_Oshanin_Blumen_1995,PRE_Lizana_Barkai_Lomholt_2010,PhysToday_Barkai_2012,JStatMech_Panja_2010,PRE_Sakaue_2013}; however, we may look at the variance $\left< \delta \Delta x(t)^2 \right>$ in the context of the driven system~\cite{PRE_Sakaue_2013,PRE_Saito_2015}.
(b) The other integral form is expressed as a convolution of the friction kernel with the difference between the response function and the velocity correlation.
The deviation from the FDR in the integral form (b) is determined to be compatible with the energetics~\cite{PRL_Harada_Sasa_2005,PRE_Harada_Sasa_2006,EPL_Speck_Seifert_2006,PRE_Deutsch_Narayan_2006}.
Note that derivations from the FDRs between (a) and (b) appear as distinct integral forms that are built by the deviation from the response function and the velocity correlation, not through incorporation of a delta-function type, but through another type kernel due to degree of non-Markovity in the polymer.\footnote{
In the presence of the step force $f(t)=f$ for $t>0$, the $q$-mode component $\left<\Lambda_q(t)\right>$ of $\left<\Lambda(t)\right>$ (eq.~(\ref{Lambda_t})) is obtained from the following integration:
\begin{eqnarray}
\frac{\left<\Lambda_q(t)\right>}{k_BT}
&-&
\frac{F_q^2}{(k_BT/N)^2}
\int_0^tdt'\int_0^tds'\, \Biggl[ \left< \frac{dX_q(t')}{dt'}\frac{dX_q(s')}{ds'} \right>\Theta(t'-s')
\nonumber \\
&&
-\frac{c_qk_BT}{N} R_q(t',s') \Biggr]
=
\frac{F_q^2}{2(k_BT/N)^2}\left< \Delta X_q(t) \right>^2.
\end{eqnarray}
Note that a part of the kernel in the double integral corresponds to the FDR deviation between the response function and velocity correlation, where the integrand does not have the kernel $\gamma_q(t')$ like eq.~(\ref{Lambda_HS}).
It should be noted that the velocity correlation is not covariance $\left< \delta \left(\frac{dX_q(t')}{dt'}\right) \delta \left(\frac{dX_q(s')}{ds'}\right) \right>$.
}

{\it ---Anomalous diffusion---}
We begin with a review of the polymer dynamics around the integral form (a).
The anomalous diffusion is defined as the temporal nonlinear power-law growth of, e.g., the MSD  $\left< \Delta x(t)^2 \right>\sim t^\alpha$ ($\alpha \neq 1$)~\cite{JChemPhys_Schiessel_Oshanin_Blumen_1995,PRE_Lizana_Barkai_Lomholt_2010,PhysToday_Barkai_2012,JStatMech_Panja_2010,PRE_Sakaue_2013}.
A formalism with the generalized Langevin equation (GLE) facilitates understanding.
Unless otherwise indicated, the forced $N$-th monomer is traced and referred to as a tagged monomer. 
Solutions to eq.~(\ref{imag_EM_TDSA}) are superimposed to give the equation of motion for $x(t)\equiv x_N(t)$ as follows:
\begin{eqnarray}
\frac{dx(t)}{dt}
&=&
\int_{-\infty}^tds\,\mu(t,s) f(s)+\eta (t),
\label{like_GLE}
\end{eqnarray}
where $\eta(t)$ denotes colored noise, and $\mu(t,s)$ is a memory kernel.
The equilibrium condition assures 
time translational symmetry and the FDR of the second kind $k_BT\mu(t-s)=\left< \eta(t)\eta(s) \right>$; i.e., eq.~(\ref{like_GLE}) is reduced to the conventional GLE~\cite{JStatMech_Panja_2010,PRE_Sakaue_2013}.
The MSD absent force, $f(t)=0$, is calculated from the solutions to eq.~(\ref{imag_EM_TDSA}),
which gives the anomalous diffusion $\left< \Delta x(t)^2 \right>_{f(t)=0}\sim t^{2/z}$
that reproduces the well-known results for the Rouse model, $\left< \Delta x(t)^2 \right>_{f(t)=0}\sim t^{1/2}$~\cite{JChemPhys_Schiessel_Oshanin_Blumen_1995,PRE_Lizana_Barkai_Lomholt_2010} with $\nu=1/2$ and $z=4$.
In the presence of the step force $f(t)=f$ for $t>0$ and $f(t)=0$ for $t<0$, the variance serves as a substitute for the MSD and shows the same power law $\left< \delta\Delta x(t)^2 \right>\sim t^{2/z}$.
Furthermore, with focus on the displacement, the FDR for the integral form (a) is confirmed as $2k_BT\left<\Delta x(t)\right>=f\left< \delta \Delta x(t)^2 \right>$.
However, these are not necessary conditions for nonequilibrium polymer stretching; the SA polymer does not satisfy these conditions.
Let us then quantify the FDR deviations for the integral form (a) with the index\footnote{The stochasticity related to $\Lambda (t)$ will be addressed later.}
\begin{eqnarray}
\Lambda (t) \equiv f\Delta x(t)-\frac{f^2\delta \Delta x(t)^2}{2k_BT},
\label{Lambda_t}
\end{eqnarray}
of which the expected value becomes $\left< \Lambda (t)\right>=0$ when the FDR holds (see appendix E).
On the other hand, placing the solutions to eq.~(\ref{imag_EM_TDSA}) (or together with eqs.~(\ref{spring}) and (\ref{friction})) into eq.~(\ref{Lambda_t}), we may find a nonzero value $\left< \Lambda (t)\right>\neq 0$ in the nonequilibrium:
\begin{eqnarray}
\left< \Lambda (t) \right>
&=& \sum_{q \geq 1}c_qN
\int_0^t ds \, {\cal T}_q(t,s) \frac{d}{ds} \left( \frac{F_q^2}{2k_q(s)} \right) (h_{q,N}^\dagger)^2
\label{G_interpretation}
\\
&=&
\left\{
 \begin{array}{ll}
0\qquad \qquad \cdots(t<\tau_f)
\\
\\
-\sum_{q=q_f}^N  \frac{f^2}{4Nk_BT}
\left[ \frac{k_BT}{k_q}-\frac{k_BT}{k_q^{(f)}} \right] 
\\
\qquad \qquad \times \left[ 1-e^{-(t-\tau_f)(q/N)^2/\tau_f} \right]^2 (h_{q,N}^\dagger)^2
\\
\qquad\qquad\quad \cdots (t>\tau_f)
\end{array}
\right.
\label{2_Ft}
\end{eqnarray}
where the integral kernel is expressed as
\begin{eqnarray}
{\cal T}_q(t,s)
&=&
\left[
\int_{s}^{t} ds'\, \frac{k_q(s')}{\gamma_q(s')} e^{-\int_{s'}^t ds''\ k_q(s'')/\gamma_q(s'')} 
\right]^2
\nonumber \\
&=&
\left[
1
-
e^{-\int_{s}^t ds'\ k_q(s')/\gamma_q(s')} 
\right]^2.
\label{2_Ft_kernel}
\end{eqnarray}
Note that an integration by parts and the equipartition of energy $\left< X_q(0)^2 \right>=k_BT/(2Nk_q)$ for $q\geq 1$ at $t=0$ are applied to obtain eq.~(\ref{G_interpretation}).
Equation~(\ref{2_Ft}) is the same as eq.~(73) in ref.~\cite{PRE_Saito_2015} with the coefficients eqs.~(\ref{spring}) and (\ref{friction}), 
while in practice we can arrive at up to eqs.~(\ref{G_interpretation}) and (\ref{2_Ft_kernel}), even without specifying the time-dependent coefficients (eqs.~(\ref{spring}) and (\ref{friction})).
The analytical results $\left< \Lambda (t) \right>$ that show the FDR deviation for the SA polymer are numerically verified by molecular dynamics simulation on the plot of $2-2\left< \Lambda (t)\right>/[f\left<\Delta x(t)\right>]$~\cite{PRE_Saito_2015}, which takes a value of 2 if the FDR holds.

{\it ---Harada-Sasa equality---}
Let us next consider the integral form (b).
The FDR deviation with the integral form (b) is quantified with an energy input rate under the nonequilibrium steady state, which is referred to as the Harada-Sasa (HS) equality~\cite{PRL_Harada_Sasa_2005}.
Although the scope of application is appropriate in the steady state, we attempt to move forward along an analogous line, even towards the transient processes.
The following calculations are performed in accordance with ref.~\cite{JStatMech_Ohkuma_Ohta_2007}, which mainly investigates the GLE for the nonMarkovian process.

A response function on the mode space is defined with a functional derivative as follows:
\begin{eqnarray}
&&
R_q(t,s)\equiv \frac{\delta}{\delta F_q(s)}\left< \frac{dX_q(t)}{dt} \right>.
\label{response_defition}
\end{eqnarray}
The path integral formulation is combined in a subsequent calculation.
Ensemble averages $\left< \cdots \right>$ are taken over the distributions of the initial configurations and trajectories for each mode space.
The probability of a trajectory given $X_q(0)$ in the mode space is denoted by ${\cal P}_q [X_q(\cdot)|X_q(0)]$, which is obtained through the occurrence probability generated by a temporal sequence of noise $\{ Z_q(t')\}$, gives the following: 
\begin{eqnarray}
{\cal P}_q [Z_q(\cdot)] &\sim& \exp{\left(-{\int_0^t dt'\,} \frac{1}{4c_qk_BT\gamma_q(t')/N}Z_q(t')^2 \right)}.
\label{PDF_noise}
\end{eqnarray}
$Z_q(t')$ in eq.~(\ref{PDF_noise}) is transformed into $X_q(t')$ with eq.~(\ref{imag_EM_TDSA})~\cite{JResNatlBurStand_Furutsu_1963,ZhEkspTeorFz_Novikov_1963,ProcConf_Donsker_1964} (see short note\footnote{
Substitution of eq.~(\ref{imag_EM_TDSA}) reinterprets eq.~(\ref{PDF_noise}) as 
\begin{eqnarray}
&& 
{\cal P}_q[X_q(\cdot)|X_q(0)] 
\nonumber \\
&\sim& J\exp{}\Biggl(-{\int_{0}^{t}dt'}\, \frac{1}{4c_qk_BT\gamma_q(t')/N}  
\nonumber \\
&&
\times \left[ 
\gamma_q(t') \frac{dX_q(t')}{dt'}
+\frac{\partial {\cal E}_q(X_q,k_q(t'))}{\partial X_q}
 -F_q(t')
\right]^2 \Biggr),
\label{PDF_noise_Xq}
\end{eqnarray}
where the Jacobian is 
$J=\exp{}\left[ \int_0^tdt'\, \frac{1}{2\gamma_q(t')} \frac{\partial^2{\cal E}_q[X_q,k_q(t')]}{\partial X_q^2} \right]$.
The functional derivative of $\left< dX_q(t)/dt \right>=\int dX_q(0){\cal P}_q(X_q(0)) \int {\cal D}X_q(s')\, \dot{X}_q(s'){\cal P}_q[X_q(\cdot)|X_q(0)]$ is with respect to the force $F_q(t')$ to give eq.~(\ref{FND_mode}).
Note that $\dot{X}_q(t)\equiv dX_q(t)/dt$, and ${\cal P}_q(X_q(0))$ denotes the probability density function for the initial condition.
}) and combining a functional derivative with respect to the external force, the difference of the response function with a velocity correlation for $t>0$ is reduced to
\begin{eqnarray}
&&
{\int_{0}^{t+\epsilon_+} ds\,}
\gamma_q(s)\delta(t-s)
\nonumber \\
&&
\times
\Biggl[
\frac{2c_qk_BT}{N}
R_q(t,s)
-
2\Biggl<
\frac{dX_q(t)}{dt} \circ \frac{dX_q(s)}{ds}
\Biggr>
\Theta (t-s)
\Biggr]
\nonumber \\
&=&
\Biggl<
\frac{dX_q(t)}{dt} \circ 
\Biggl[ \frac{\partial {\cal E}_q(X_q,k_q(t))}{\partial X_q}-F_q(t) \Biggr] \Biggr>.
 \label{FND_mode}
\end{eqnarray}
Integrating over time and superimposing the mode components, we then organize for $t>0$:
\begin{eqnarray}
\left< f\Delta x(t) \right>
&=&
\left<\Lambda^{(HS)}(t)\right> +\left<\Delta {\cal E}(t)\right>
\label{mode_energy_balance}\\
&&-\sum_{q\geq 1}c_qN\int_0^tdt'\,\left< \frac{\partial {\cal E}_q^{(f)}}{\partial t}\right>  (h_{q,N}^\dagger)^2,
\nonumber 
\end{eqnarray}
where the first term on the right side of eq.~(\ref{mode_energy_balance}) is given by
\begin{eqnarray}
\left<\Lambda^{(HS)}(t)\right>
&\equiv&
\sum_{q}c_qN\int_0^tdt'\,
\gamma_q(t')
\Biggl[ \left<\frac{dX_q(t')}{dt'} \circ \frac{dX_q(t')}{dt'}\right>  
\nonumber \\
&&
-\frac{2c_qk_BT}{N} R_q(t',t') \Biggr](h_{q,N}^\dagger)^2.
\label{Lambda_HS}
\end{eqnarray}
The HS equality focuses on $\left<\Lambda^{(HS)}(t)\right>$, which consists of the difference between the velocity correlation and the response functions.
Furthermore, the extra terms on the right-hand side of eq.~(\ref{mode_energy_balance}) arise as a consequence of the chainlike internal degrees of freedom, i.e., the global polymer configurations.
The stretching polymer eventually settles into the steady state with finite variance $\left< X_q(t)^2 \right>$, which is eliminated when taking a long time limit in the time average of the second term of eq.~(\ref{mode_energy_balance}).
This indicates $\lim_{t\rightarrow +\infty} \langle \Delta {\cal E}(t) \rangle /t=0$ and 
$\lim_{t\rightarrow +\infty}  (1/t)\int_0^tdt'\, (\partial \langle {\cal E}_q^{(f)}(t')\rangle /\partial t') =0$.
We thus arrive at $\lim_{t\rightarrow \infty}[\left<f \Delta x(t)\right> -\left<\Lambda^{(HS)}(t)\right>]/t=0$.
This means that the energy input rate $\left< f\Delta x(t)\right>/t$ that corresponds to the rate of genuine mechanical work for the present system is balanced with $\left<\Lambda^{(HS)}(t)\right>/t$ for $t\rightarrow +\infty$, as expressed by the HS equality that quantifies the FDR violation with the energy input rate.

Equation~(\ref{mode_energy_balance}) is rewritten as $ \langle W\rangle=\langle\Lambda^{(HS)}\rangle+\langle \Delta {\cal E}^{(f)} \rangle$ at time $t$ with eq.~(\ref{fic_work}),\,(\ref{SP_Hq}). 
Considering the transient process carefully, 
we identify $\left<\Lambda^{(HS)}(t)\right>$ by comparison with an ensemble average of eq.~(\ref{energy_conservation_law}) as
\begin{eqnarray}
\bigl<\Lambda^{(HS)}(t)\bigr>=\left<Q(t)\right>.
\label{enr_conv_ave}
\end{eqnarray}
When $q$-mode components are represented with the subscript $q$, 
the approximated description (or the rigorous description for the Rouse polymer) also satisfies each energy balance $\langle W_q\rangle=\langle \Lambda_q^{(HS)}\rangle+\langle \Delta {\cal E}_q^{(f)} \rangle$ with the analogous interpretation $\langle \Lambda_q^{(HS)}(t)\rangle=\langle Q_q(t)\rangle$.
Thus, $\langle \Lambda^{HS}(t) \rangle$ or $\langle \Lambda_q^{HS}(t) \rangle$ is identified as the mean heat within the present definition of the work and the effective Hamiltonian.

Recall the observation of $\left< \Lambda (t) \right>$ (eqs.~(\ref{G_interpretation}) and (\ref{2_Ft})) with the integral form (a) in order to see an association with the integral form (b).
Here, we directly compare eq.~(\ref{mode_energy_balance}) with $\left<\Lambda(t)\right>$ by the introduction of an elasticity-related term as
\begin{eqnarray}
\Lambda^{(ela)}(t)
&\equiv&
\sum_{q\geq 1}
\biggl[\int_0^tdt'\,c_qN\frac{\partial {\cal E}_q}{\partial X_q}\circ \frac{dX_q(t')}{dt'}
\nonumber \\
&&
-\frac{N^2F_q^2}{2k_BT} \delta \Delta X_q(t)^2 \biggr](h_{q,N}^\dagger)^2,
\label{Lambda_ela}
\end{eqnarray}
where the second term in the square brackets on the right-hand side represents the mode components decomposed from the quadratic term in eq.~(\ref{Lambda_t}), i.e., $f^2\left< \delta\Delta x(t)^2\right>=\sum_q N^2F_q^2 \left< \delta \Delta X_q(t)^2 \right>(h_{q,N}^\dagger)^2$, of which the terminal variance $\left< \delta \Delta X_q(+\infty)^2 \right>=k_BT/(2Nk_q(+\infty))$ is written with the equipartition of energy as the static quantities (see appendix E).
We then organize the FDR deviations as
\begin{eqnarray}
\left<\Lambda(t)\right>
&=&
\left<\Lambda^{(HS)}(t)\right> +\left<\Lambda^{(ela)}(t)\right>
\label{FDR_dev_comp}
\end{eqnarray}
where
\begin{eqnarray}
\left<\Lambda^{(HS)}(t)\right>
&=&
\int dn'\int dn''\, \gamma_{n'-n''}(t)
\left< \dot{x}_{n'}(t)\dot{x}_{n''}(t) \right>
\nonumber \\
&&
-2k_BT\int dn'\, \gamma_{N-n'}(t) \frac{\delta \left< \dot{x}_N(t) \right>}{\delta f_N(t)}
\nonumber \\
\left<\Lambda^{(ela)}(t)\right>
&=&
\int dn'\int dn''\int_0^tdt'\, 
\nonumber \\
&&
\times \frac{\kappa_{n'-n''}(t')}{2} 
\frac{d}{dt'} \left< x_{n'}(t')x_{n''}(t') \right>
\nonumber \\
&&
-\frac{f^2\left< \delta\Delta x(t)^2\right>}{2k_BT},
\label{FDR_dev_comp_real}
\end{eqnarray}
in which $\kappa_{n-n'}(t)\equiv \sum_{q=1}^N c_q^{-1}Nk_qh_{q,n} h_{q,n'}$ denotes the spring kernel in real space.
An interesting point is that $\left< \Lambda^{(HS)}(t)\right>$ is in contrast to $\left<\Lambda^{(ela)}(t)\right>$ because the former and latter are closely related to the dynamical and static quantities, respectively.
The former, as one of the cases, even includes the hydrodynamic long-range interaction, in which the dissipation mechanism is incorporated as the convolution of the power-law-decaying frictional kernel $\gamma_{n'-n''}(t)$.
The latter has the convolution of a spring kernel $\kappa_{n'-n''}(t)$ with the monomer indices $n'$ and $n''$, which is derived from the effective Hamiltonian based on the mode description.
Thus, the difference between $\left< \Lambda(t)\right>$ and $\left< \Lambda^{(HS)}(t)\right>$, i.e., between the integral forms (a) and (b), is $\left<\Lambda^{(ela)}(t)\right>$, which is closely associated with the static quantity.
The terminal difference is thus given only by the static quantities.

As another significant point, an analogue to the Bochkov-Kuzovlev relation could possibly be introduced.
The ensemble average $\left< \cdot \right>$ taken over the preaveraged Gaussian distribution gives a simple expression:
\begin{eqnarray}
\left< e^{-f\Delta x(t)/k_BT} \right> = e^{-\left< \Lambda(t) \right>/k_BT},
\label{ave_exp_fx}
\end{eqnarray}
whereas a non-Gaussian distribution before preaveraging or addition of the fictive part of the mechanical work provides the well-known Bochkov-Kuzovlev relation that corresponds to $\left< \Lambda(t) \right>=0$.
Recall that $\left< \Lambda(t) \right>$ equals the FDR deviations.
Incidentally, if $\left< \Lambda(t) \right>$ on the right-hand side were viewed as a thermodynamic function defined with eq.~(\ref{ave_exp_fx}), then a downward concavity of the exponential mathematically would lead to the following inequality:
\begin{eqnarray}
\left< f\Delta x(t) \right> \geq -k_BT\log{e^{-\left< \Lambda(t) \right>/k_BT}},
\end{eqnarray}
where the derivation calculus appears similar to that of the second law of thermodynamics $W\geq \Delta F$ from $\left< e^{-W/k_BT} \right> = e^{-\Delta F/k_BT}$ with $\Delta F$ denoting the free energy difference.
As in $e^{-\Delta F/k_BT}$, if $e^{-\left< \Lambda(t) \right>/k_BT}$ is considered as a ratio of the partition-function-like quantities, then $e^{-\left< \Lambda(t) \right>/k_BT}$ corresponds to a ratio of deviations of normalizations in the Gaussian distributions due to the $-fx$ term.\footnote{
Explicitly, we may write
\begin{eqnarray}
e^{-\left< \Lambda(t) \right>/k_BT}
=
\prod_q 
\frac{
\int dX_q\, \frac{1}{\sqrt{2\pi c_qk_BT/(Nk_q)}} \exp{\left[-\frac{k_qX_q^2/2 -F_qX_q}{c_qk_BT/N}\right]}
}
{
\int dX_q\, \frac{1}{\sqrt{2\pi c_qk_BT/(Nk_q^{(f)})}} \exp{\left[-\frac{k_q^{(f)}X_q^2/2 -F_qX_q}{c_qk_BT/N}\right]}
}.
\nonumber \\
\end{eqnarray}
In the absence of the $-fx$ term, i.e., $F_q=0$, the Gaussian distributions of the integrands in both the denominator and the numerator are normalized, and then we find $\left< \Lambda(t) \right>=0$.
}

\section{Discussion}
\label{discussion}

Section~\ref{FDR_dev} has discussed the ensemble average quantities.
Let us consider eq.~(\ref{enr_conv_ave}).
The equality emerging on the average statistics suggests to look for $\Lambda^{(HS)}(t)= Q(t)$ at the stochastic level.
If an instantaneous stochastic response function in the mode space is introduced for the Langevin dynamics with 
\begin{eqnarray}
R^*_q(t,t)= -k_B^{-1}\frac{\delta [-k_B\log{{\cal P}_q(X_q(t))}]}{\delta F_q(t)} \circ \frac{dX_q(t)}{dt},
\label{stochastic_response}
\end{eqnarray}
then  under Gaussian white noise, we may define
\begin{eqnarray}
\Lambda^{(HS)}(t)
&\equiv&
\sum_q c_qN
\int_0^tdt'\,
\gamma_q(t')
\Biggl[ 
\nonumber \\
&&
\frac{dX_q(t')}{dt'} \circ \frac{dX_q(t')}{dt'}
-\frac{2c_qk_BT}{N}R_q^*(t',t')\Biggr](h_{q,N}^\dagger)^2.
\nonumber \\
\label{stochastic_HS_def}
\end{eqnarray}
such that $\Lambda^{(HS)}(t)= Q(t)$.
Calculation of $R^*_q(t,t)$ (see appendix F) and comparison with eq.~(\ref{SP_Qq}) reveals equivalence under the Gaussian white noise:
\begin{eqnarray}
\Lambda^{(HS)}_q(t) =Q_q(t).
\label{Q_LambdaHS}
\end{eqnarray}
We confirm that $\left<R^*_q(t,t)\right>=R_q(t,t)$.
In addition, the function $s_q(t)=-k_B\log{{\cal P}_q(X_q(t))}$ that appears in eq.~(\ref{stochastic_response}) corresponds to the stochastic entropy~\cite{RepProgPhys_Seifert_2012,EPL_Seifert_Speck_2010}. 
If the equivalent time quantities (e.g., $t=s$ in eq.~(\ref{response_defition})) are focused on exclusively,
then the ensemble average level with $R_q(t,t)=\left<R^*_q(t,t)\right>$ leads us to a special case of the FDR expression with the stochastic entropy~\cite{RepProgPhys_Seifert_2012,EPL_Seifert_Speck_2010}.

To reveal the FDR deviations, this study has so far dealt with an SA polymer, of which the effective interaction range is nonlocal.
However, it is stressed that the nonlocal interaction is not a necessary condition to cause the FDR deviations.
For example, let us modify the Rouse model such that the elastic force varies with time, such as $f_n^{(ela)}=k(t)\partial^2 x_n/\partial n^2$, while the range is kept local (the second derivative with respect to $n$ indicates the local interaction).
However, analogous arguments can be made, and then we find the FDR deviations $\Lambda(t)\neq 0$.
In practice, temporal changes in the stiffness could be widely observed, probably in intracell situations.

\section{Concluding remarks}
\label{conclusion}

This article has discussed the preaveraging description of the temporal evolution of the regime structure for polymer stretching in terms of the nonequilibrium work relation and the FDR deviations.
The effective spring constants are considered as spontaneously varying in the sense of the physical origin.
However, the preaveraging hides the spontaneous properties of the elemental degrees of freedom that build up the effective coefficients.
In applying the preaveraging description, consistency may require the effective spring constants to be interpreted as fictive external parameters.
In the present mode formalism, the interactions from different modes or from the other structure regimes are given via the parameters (the spring constants and the friction coefficients) based on preaveraging, and the respective regime levels are rather viewed as distinct systems independent of each other.
Therefore, 
a change in the effective Hamiltonian may be considered as an energy transfer from the exterior, even if it is actually from the interior in light of the entire regime system.

Significant developments in visualization techniques in cells have placed more demands in terms of interpretations from physics~\cite{PhysToday_Barkai_2012}.
This study has considered a significantly simplified nonequilibrium polymer system compared to biological systems.
Nonetheless, we hope that the present study would be helpful for cell studies.
For example, refs.~\cite{EPL_Grosberg_1993,Science_LiebermanAiden_2009} pointed out the importance of the global regime often called crumpled or fractal globules.
It would not be very surprising if those hierarchal regime structures were not temporally eternal, but evolutional, according to nonequilibrium processes in metabolism.

\section*{Acknowledgement}
The author thanks T. Sakaue for fruitful discussions and critical reading.

\section*{Appendix}

\subsection{Rouse polymer}

A Rouse model illustrates a rigorous conversion between the real and the mode coordinates.
The equation of motion in real space for the discrete model is governed by
\begin{eqnarray}
\gamma \frac{\partial x_n(t)}{\partial t}
&=&
f_n^{(ela)}(t) +f_n(t) +\zeta_n(t),
\label{Rouse_EM_real}
\end{eqnarray}
where $f_n^{(ela)}=-\partial {\cal E}/\partial x_n$ denotes the elastic force produced by harmonic potentials ${\cal E}=(k/2)\sum_{n'=0}^{N-1} (x_{n'+1}-x_{n'})^2$.
Taking the continuum limit, we see that $f_n^{(ela)} = k(x_{n+1}-2x_{n}+x_{n-1}) \rightarrow k\partial^2 x_n/\partial n^2$.
Unless otherwise noted, the main text in the article employs the continuum picture.

The integer exponent is ``2" in the spring constant $k_q=k(q/N)^2$ with $\nu=1/2$ in the mode space for the Rouse polymer, which is followed by the second-order derivative $-k\partial^2 x_n/\partial n^2$ in real space.
The integer exponent indicates the local interaction.
On the other hand, the SA effects or the hydrodynamic interactions (HIs) are long-range interactions, which are represented by, roughly speaking, fractional derivatives with non-integer exponent $\nu$ in real space.

\subsection{Difference in thermodynamic potential}
\label{cal_1}

A difference in the thermodynamic potential for fixed-tension ensemble is calculated from eqs.~(\ref{SP_Uq}) and (\ref{SP_Hq}). 
For $t \gg \tau$, $\Phi (t)$ is obtained as 
\begin{eqnarray}
\Phi (t)
&=& -\sum_{q\geq 1} [c_qN(h_{q,N}^\dagger)^2] \frac{c_qk_BT}{N} \log{}\Biggl[ \int_{-\infty}^{+\infty}dX_q\,
\nonumber \\
&& \qquad \qquad \times \exp{\left(-\frac{{\cal E}_q^{(f)}}{(c_qk_BT/N)}  \right)} \Biggr]
\nonumber \\
&=& -\sum_{q\geq 1} 
[c_qN(h_{q,N}^\dagger)^2]
\frac{c_qk_BT}{N}
\log{} \Biggl[ \sqrt{\frac{2\pi c_qk_BT}{Nk_q^{(f)}}} 
\nonumber \\
&& \qquad \qquad \qquad \times \exp{\left( \frac{F_q^2}{2c_qk_BTk_q^{(f)}/N}\right)} \Biggr] 
\nonumber \\
&=& 
\sum_{q\geq 1} \left[ 
-\frac{F_q^2}{4(k_q^{(f)}/N)}
-\frac{k_BT}{4} \log{\sqrt{\frac{\pi k_BT}{Nk_q^{(f)}}} }
\right]
(h_{q,N}^\dagger)^2.
\nonumber \\
\end{eqnarray}
with ${\cal E}_q^{(f)}(X_q,k_q(t),F_q(t))$.
Also, for $t=0$, we get
\begin{eqnarray}
\Phi (0)
&=& 
-\frac{k_BT}{4} \sum_{q\geq 1} \log{\sqrt{\frac{\pi k_BT}{Nk_q}}} (h_{q,N}^\dagger)^2.
\end{eqnarray}
From $\Delta \Phi=\Phi (t)-\Phi (0)$, we then arrive at eq.~(\ref{F_Wima}).

For the Rouse model, the left-and side of eq.~(\ref{fic_W_F}) is easily calculated as follows:
\begin{eqnarray}
&&\left< e^{-W_{q\geq 1}/(k_BT)} \right>
\nonumber \\
&=&\left< \exp{ \Biggl[ -\sum_q N\frac{F_q(\partial {\cal E}_q^{(f)}/\partial F_q)|_{t=0}}{k_BT}c_q(h_{q,N}^\dagger)^2 \Biggr]} \right>
\nonumber \\
&=&\left< \exp{ \Biggl[ \sum_q N\frac{F_qX_q(0)}{k_BT}c_q(h_{q,N}^\dagger)^2 \Biggr] } \right>
\nonumber \\
&=&\exp{ \Biggl[ \sum_{q\geq 1} N^2\frac{F_q^2\left< X_q(0)^2\right>}{2(k_BT)^2}c_q^2(h_{q,N}^\dagger)^4  \Biggr] } 
\nonumber \\
&=&\exp{ \Biggl[ \sum_{q \geq 1} \frac{f^2}{4Nk_q}(h_{q,N}^\dagger)^2  \Biggr] }, 
\end{eqnarray}
where $c_q h_{q,N}^\dagger=(-1)^q$ and $(\partial {\cal E}_q^{(f)}/\partial F_q)|_{t=0}\equiv \int_0^tdt'\,F_q \delta (t'-\epsilon_+)$ are used.
In the case of the Rouse polymer with the invariant coefficients $k_q(t)=k_q$ and $\gamma_q(t)=\gamma_q$, only the first term in brackets on the right-hand side of eq.~(\ref{F_Wima}) survives.

\subsection{Energy balance modified by preaveraging}
\label{EB_preave}

We consider energy balance modified by preaveraging.
There are two descriptions for the Langevin equations: (i) an elementary expression before preaveraging, and (ii) that after preaveraging.
While the latter is extensively discussed in the main text (e.g., see eq.~(\ref{imag_EM_TDSA}) with eq.~(\ref{SP_Uq})), the elementary equation of motion for the former is represented by eq.~(\ref{real_original_EM}).
The elementary effective Hamiltonian before preaveraging is explicitly expressed by
\begin{eqnarray}
\nonumber \\
{\cal E}_{(e)}^{(f)}
&=&
{\cal E}_{(e)}-fx_N
\\
{\cal E}_{(e)}
&=&
\sum_n \frac{1}{2}k(x_{n+1}-x_n)^2
+
\sum_{i>j} U_{SA}(|x_i-x_j|),
\end{eqnarray}
where $U_{SA}(|x_i-x_j|)$ represents the SA interaction potential similar to the Lennard-Jones potentials, hard core interaction potentials, or a corresponding part $(v_{SA}/2)\int_0^Ndn\int_0^Ndn'\,\delta (x_n-x_{n'})$ of the Edwards Hamiltonian with $v_{SA}$ being the SA interaction parameter~\cite{PhysRep_Vilgis_2000}.
The energy balance for the respective Langevin equations is maintained as
\begin{eqnarray}
d{\cal E}_{(e)}^{(f)} &=& d'W_{(e)}-d'Q_{(e)}
\label{energy_balance_elementary}
\\
d{\cal E}^{(f)} &=& d'W-d'Q.
\label{energy_balance_repeated}
\end{eqnarray}
Note that eq.~(\ref{energy_balance_repeated}) is repeated and essentially the same as eq.~(\ref{energy_conservation_law}).

According to Sekimoto's definition of heat~\cite{SekimotoBook}, we have\begin{eqnarray}
d'Q_{(e)}
&=&
\sum_n\left(
\sum_{n'}
\Gamma [ x_n,x_{n'} ] \frac{dx_{n'}}{dt}
-\zeta_n(t)
\right)\circ dx_n.
\end{eqnarray}
We require the condition that the elementary heat changes into the preaveraged heat as
\begin{eqnarray}
\hat{\mathrm{P}} \left( d'Q_{(e)}\right)=d'Q,
\label{heat_preaveraging}
\end{eqnarray}
where the preaveraging operator is introduced with $\hat{\mathrm{P}}(\cdot)=\sum_q \hat{\mathrm{P}}_q\left( \cdot \right)$.
Let $\hat{\mathrm{P}}_q\left( \cdot \right)$ be the preaveraging operator extracting the $q$-mode defined as
\begin{eqnarray}
\hat{\mathrm{P}}_q \biggl( \cdot \biggr)
&=&
\int \prod_{q' \neq q} dX_{q'}\, \rho_q(\{ X_{q'}(t) \})\,\biggl( \cdot \biggr),
\end{eqnarray}
where
\begin{eqnarray}
\rho_q(\{ X_{q'}(t) \})
\equiv
\frac{ {\cal P}_{(e)} (\{ X_{q'}(t) \}) }
{ \int \prod_{q'' \neq q} dX_{q''}\,{\cal P}_{(e)} (\{ X_{q''}(t) \}) }
\label{rho}
\end{eqnarray}
and 
\begin{eqnarray}
{\cal P}_{(e)} (\{ X_{q}(t) \}) 
&=&
\int \prod_{q''} {\cal D}X_{q''}\, dX_{q''}(0)
\nonumber \\
&&
\times {\cal P}_{(e)}[\{X_{q'}(\cdot)|X_{q'}(0)\}] {\cal P}_{(e)}(\{X_{q'}(0)\}).
\end{eqnarray}
Here ${\cal P}_{(e)}[\{X_{q'}(\cdot)|X_{q'}(0)\}]$ denotes the probability of the path from $\{X_{q'}(0)\}=\{X_0(0),X_1(0),\cdots,X_N(0)\}$ to $\{X_{q'}(t)\}=\{X_0(t),X_1(t),\cdots,X_N(t)\}$, and ${\cal P}_{(e)}(\{X_{q'}(0)\})$ is the probability density of the initial condition.
Keep in mind that eq.~(\ref{rho}) is normalized to unity for each mode $q$:
\begin{eqnarray}
\int \prod_{q'\neq q}dX_{q'}\,{\rho}_{q} (\{ X_{q'} \})  =1.
\end{eqnarray}
The operator does not essentially alter the dissipation for a free-draining polymer, because $\hat{\mathrm{P}} \left( d'Q_{(e)}\right)=d'Q_{(e)}=d'Q$ exactly, whereas the nonlocal kernel $\Gamma [x_n,x_m ]$ with the relative monomer position for the non-draining polymer is reduced to $\Gamma (n-m)$ with the relative monomer index.
Equation~(\ref{heat_preaveraging}) is the assumption of the preaveraging, and the noise with the viscous friction is transformed such that the FDR of the second kind is maintained (refer to the supplement in appendix D).

In addition, a condition is imposed on the effective Hamiltonian as
\begin{eqnarray}
\hat{\mathrm{P}} \left( {\cal E}_{(e)}^{(f)}\right)={\cal E}^{(f)},
\label{Ee_E}
\end{eqnarray}
which also indicates that the value of the effective Hamiltonian is maintained after preaveraging.

We move on to observation of an infinitesimal.
A key point is that $\hat{\mathrm{P}} \left( d'Q_{(e)}\right)=d'Q$ and $\hat{\mathrm{P}} \left( {\cal E}_{(e)}^{(f)}\right)={\cal E}^{(f)}$; however, generally $\hat{\mathrm{P}} \left( d{\cal E}_{(e)}^{(f)}\right)\neq d{\cal E}^{(f)}$.
A change in the effective Hamiltonian by preaveraging is obtained as
\begin{eqnarray}
d{\cal E}^{(f)}
&=& 
\hat{\mathrm{P}}\left( d{\cal E}_{(e)}^{(f)}\right)
\nonumber \\
&&
+
\sum_q \left[ \int \prod_{q' \neq q} dX_{q'}\, {\cal E}_{(e)}^{(f)}(\{ X_{q'} \}) d\rho_q (\{ X_{q'} \})\right],
\nonumber \\
\label{der_dEe_dE}
\end{eqnarray}
where 
\begin{eqnarray}
\hat{\mathrm{P}}\left( d{\cal E}_{(e)}^{(f)}\right)
=
\sum_q \int \prod_{q' \neq q} dX_{q'}\,\rho_q (\{ X_{q'} \})
d{\cal E}_{(e)}^{(f)}(\{ X_{q'} \})
\nonumber \\
\end{eqnarray}
If $d\rho_q(\{ X_{q'} \})=0$ holds, then 
eq.~(\ref{der_dEe_dE}) indicates $d{\cal E}^{(f)}=\hat{\mathrm{P}} \left( d{\cal E}_{(e)}^{(f)}\right)$; however, generally $d\rho_q (\{ X_{q'} \})=\sum_q (\partial \rho_q/\partial X_q)dX_q+(\partial \rho_q/\partial t)dt\neq 0$.
Application of the preaveraging operator into eq.~(\ref{energy_balance_elementary}) gives
\begin{eqnarray}
\hat{\mathrm{P}}\left( d{\cal E}_{(e)}^{(f)}\right) 
=\hat{\mathrm{P}}\left( d'W_{(e)}\right)-\hat{\mathrm{P}}\left( d'Q_{(e)}\right)
\end{eqnarray}
where $\hat{\mathrm{P}}\left( d'W_{(e)}-d'Q_{(e)}\right)=\hat{\mathrm{P}}\left( d'W_{(e)}\right)-\hat{\mathrm{P}}\left( d'Q_{(e)}\right)$ is followed from the linearity of $\hat{\mathrm{P}}(\cdot)$.
Recalling eqs.~(\ref{heat_preaveraging}) and (\ref{der_dEe_dE}), we find
\begin{eqnarray}
d'W
&=&
\hat{\mathrm{P}}\left( d'W_{(e)}\right)
\nonumber \\
&&
+
\sum_q \left[ \int \prod_{q' \neq q} dX_{q'}\, {\cal E}_{(e)}^{(f)}(\{ X_{q'} \}) d\rho_q (\{ X_{q'} \}) \right].
\nonumber \\
\end{eqnarray}
Thus, the work may also be modified via preaveraging.

The linearity of the term with the ``genuine" external parameter $f(t)x_N=f(t)\sum_q X_qh_{q,N}^\dagger$ indicates $\hat{\mathrm{P}}\left( f(t)x_N\right)=f(t)x_N$ under a transformation with eq.~(\ref{rho}).
Thus, in the discussion on how the fictive external parameters emerge, the genuine external parameter $f(t)$ is separated from the effective Hamiltonian, as in 
\begin{eqnarray}
{\cal E}_{(e)}(\{ X_{q} \})={\cal E}_{(e)}^{(f)}(\{ X_{q} \},f)+f\sum_q X_qh_{q,N}^\dagger.
\end{eqnarray}
Defined as ${\cal E}_{(e)}(\{ X_{q} \})$, which does not have the external parameters, an infinitesimal in the effective Hamiltonian is generally given by
\begin{eqnarray}
d{\cal E}_{(e)}
=
\sum_q\frac{\partial {\cal E}_{(e)}}{\partial X_{q}}\circ dX_q,
\label{Ee_CoarseGraining}
\end{eqnarray}
and the preaveraged infinitesimal is formally written as
\begin{eqnarray}
d{\cal E}
=
\sum_q\frac{\partial {\cal E}}{\partial X_{q}}\circ dX_q
+
\sum_{\lambda_q=k_q} 
\frac{\partial {\cal E}}{\partial \lambda_q} \dot{\lambda}_q dt,
\label{E_CoarseGraining}
\end{eqnarray}
where $\{\lambda_q\}=\{k_0,k_1,k_2,\cdots,k_N\}$ represents a set of the ``fictive" external parameters that emerge after preaveraging.
The force balance is assumed to be sustained after the preaveraging, i.e., $-\partial {\cal E}_{(e)}/\partial X_{q}$ is converted to $-\partial {\cal E}/\partial X_{q}$ from the elementary to the preaveraged levels, and then
\begin{eqnarray}
-\sum_q\frac{\partial {\cal E}}{\partial X_{q}}\circ dX_q
=
\hat{\mathrm{P}}\left( -\sum_q\frac{\partial {\cal E}_{(e)}}{\partial X_{q}}\circ dX_q \right).
\end{eqnarray}
Comparison of the preaveraged eq.~(\ref{Ee_CoarseGraining}) with eq.~(\ref{E_CoarseGraining}) indicates
\begin{eqnarray}
d{\cal E}
&=&
\hat{\mathrm{P}} \left( d{\cal E}_{(e)} \right)
+
\sum_{\lambda_q=k_q} 
\frac{\partial {\cal E}}{\partial \lambda_q} d\lambda_q
\label{dE_dEe_lambda}
\end{eqnarray}
Recalling eq.~(\ref{der_dEe_dE}), and comparing it with eq.~(\ref{dE_dEe_lambda}), the additional term is identified as
\begin{eqnarray}
\frac{\partial {\cal E}}{\partial \lambda_q} d\lambda_q
=
\int \prod_{q' \neq q} dX_{q'}\, {\cal E}_{(e)}(\{ X_{q'} \}) d\rho_q (\{ X_{q'} \}),
\end{eqnarray}
which means that $d\rho_q (\{ X_{q'} \})$ becomes the time-dependent term mediated by the fictive external parameters.

\subsection{Heat modified by preaveraging}

We see the preaveraging assumption regarding the heat from a viewpoint of the Crooks fluctuation theorem (FT)~\cite{PRE_Crooks_1999}.
The Crooks FT for the elementary Langevin dynamics from initial time $0$ to $t$ is written as
\begin{eqnarray}
\frac{{\cal P}_{(e)}[\{X_q(\cdot)|X_q(0)\};f]}{{\cal P}_{(e)}[\{X_q^\dagger(\cdot)|X_q^\dagger(0)\};f^\dagger]}
=
e^{Q_{(e)}/k_BT},
\label{elementary_Crooks}
\end{eqnarray}
where 
${\cal P}_{(e)}[\{X_q(\cdot)|X_q(0)\};f]$ denotes the probability of the path from $\{X_q(0)\}=\{X_1(0),\,X_2(0),\,\cdots,X_N(0)\}$ to $\{X_q(t)=\{X_1(t),\,X_2(t),\,\cdots,X_N(t)\}\}$ under the protocol with the external parameter $f(t)$, and ${\cal P}_{(e)}[\{X_q^\dagger(\cdot)|X_q^\dagger(0)\};f]$ denotes the probability of the reverse path from $\{X_q^\dagger(0)\}=\{X_q(t)\}$ to $\{X_q^\dagger(t)\}=\{X_q(0)\}$ under the reverse protocol $f^\dagger (t')=f(t-t')$ for $0\leq t' \leq t$.
$Q_{(e)}$ denotes heat that satisfies the first law of thermodynamics, $\Delta {\cal E}_{(e)}^{(f)}=W_{(e)}-Q_{(e)}$.
Note that in the present notation, the external parameter $f(t)$ explicitly appears in the argument of the probability of path.

In an analogous way, we expect the FT for the preaveraged Langevin dynamics:
\begin{eqnarray}
\prod_q
\frac{{\cal P}_q[\{X_q(\cdot)\}|X_q(0); \Lambda_q]}{{\cal P}_q[\{X_q^\dagger(\cdot)\}|X_q^\dagger(0); \Lambda_q^\dagger]}
=
e^{Q/k_BT},
\label{preave_Crooks}
\end{eqnarray}
where $\Lambda_q\equiv( \lambda_q,f)$ includes both the fictive and genuine external parameters, ${\cal P}_q[\{X_q(\cdot)\}|X_q(0); \Lambda_q]$ denotes the probability of the path from $X_q(0)$ to $X_q(t)$ on $q$-mode, and the probability of the reverse path is represented by ${\cal P}_q[X_q^\dagger(\cdot)|X_q^\dagger(0); \Lambda^\dagger_q]$.
The first law of thermodynamics is transformed as $\Delta {\cal E}^{(f)}=W-Q$.

The preaveraging assumption $\hat{\mathrm{P}}(Q_{(e)})=Q$ means
\begin{eqnarray}
\hat{\mathrm{P}}_q
\left(
\log{
\frac{ {\cal P}_{(e)}[\{X_q(\cdot)|X_q(0)\};f]}{{\cal P}_{(e)}[\{X_q^\dagger(\cdot)|X_q^\dagger(0)\};f^\dagger]}
}
\right)
&=&
\log{
\frac{ {\cal P}_q[X_q(\cdot)|X_q(0); \Lambda_q] }{ {\cal P}_q[X_q^\dagger(\cdot)|X_q^\dagger(0); \Lambda^\dagger_q] }
}.
\nonumber \\
\label{preave_Prob_ratio}
\end{eqnarray}
Equation~(\ref{preave_Prob_ratio}) holds if the preaveraging is performed so as to maintain
\begin{eqnarray}
&&
\hat{\mathrm{P}}_q
\left( \left[ \sum_{q'}\Gamma[X_q,\{ X_{q'}\}] \frac{dX_{q'}}{dt} -Z_{(e),q}(t)\right] \circ \frac{dX_{q}}{dt} \right)
\nonumber \\
&=&
\left[ \gamma_q \frac{dX_{q}}{dt} -Z_q(t)\right] \circ \frac{dX_{q}}{dt}.
\label{preave_Prob_ratio_2}
\end{eqnarray}
and the FDR of the second kind describes the noise distribution ${\cal P}_q[\{ Z_q(\cdot)\}]\sim \exp{(-\int_0^t dt'\,Z_q(t')^2/(2\gamma_qk_BT))}$.
That is, if the preaveraging satisfies eq.~(\ref{preave_Prob_ratio_2}) and the FDR of the second kind is under Gaussian white noise, then we find the FT for the preaveraged description (eq.~(\ref{preave_Crooks})).
Note that $Z_{(e),q}(t)$, and $\Gamma[X_q,\{ X_{q'}\}]$ are formal expressions before preaveraging.

\subsection{FDR deviation}
\label{cal_mode_FDR}

Provided that the system is in equilibrium, the FDR for each mode holds: 
\begin{eqnarray}
F_q\left< \Delta X_q(t) \right>&=& F_q^2\frac{\left< \delta \Delta X_q(t)^2 \right>}{2(c_qk_BT/N)}.
\label{FDR_mode_app}
\end{eqnarray}
with the external force in the mode space being given by $F_q=(-1)^qf/N$.
Bear in mind that the FDR in the real space is written as $\left< \Lambda (t) \right>=0$.

Expansion of the cumulant facilitates understanding the analytical structure by using a property of the Gaussian process.
In the real space, we know that a characteristic function is written with first and second moments:
\begin{eqnarray}
\left< \exp{\left[-\frac{f\Delta x(t)}{k_BT}\right]} \right> 
&=&
\exp{\left(-\frac{f\left< \Delta x(t)\right>}{k_BT} +\frac{f^2\left< \delta \Delta x(t)^2\right>}{2(k_BT)^2} \right)}.
\nonumber \\
\label{cumulant_ex_1}
\end{eqnarray}
The conversion rule rewrites the argument of the exponential on the left-hand side as
\begin{eqnarray}
\frac{f\Delta x(t)}{k_BT}
&=&\sum_{q} Nc_q\frac{F_q\Delta X_q(t)}{k_BT} (h_{q,N}^\dagger)^2
\nonumber \\
&=&\sum_{q} \frac{(-1)^qc_qF_q\Delta X_q(t)}{c_qk_BT/N}h_{q,N}^\dagger,
\end{eqnarray}
where $c_q h_{q,N}^\dagger=(-1)^q$ is used.
Then, eq.~(\ref{cumulant_ex_1}) is calculated in the mode space.
\begin{eqnarray}
\left< \exp{\left[-\frac{f\Delta x(t)}{k_BT}\right]} \right> 
&=&
\left< \exp{\left[-\frac{\sum_{q} (-1)^qc_qF_q\Delta X_q(t)}{(c_qk_BT/N)}h_{q,N}^\dagger\right]} \right>
\nonumber \\
&=&
\prod_q \left< \exp{\left[-\frac{{(-1)^qc_qF_q\Delta X_q(t)}}{c_qk_BT/N} h_{q,N}^\dagger \right]} \right>
\nonumber \\
&=&
\exp{}\Biggl( \sum_{q=0}^N \Biggl[ -\frac{{(-1)^qc_qF_q\left< \Delta X_q(t)\right> h_{q,N}^\dagger}}{(c_qk_BT/N)} 
\nonumber \\
&&
+\frac{1}{2}\frac{{(c_qF_q)^2\left< \delta \Delta X_q(t)^2\right> (h_{q,N}^\dagger)^2}}{(c_qk_BT/N)^2}  \Biggr] \Biggr).
\label{cumulant_ex_2}
\end{eqnarray}
Inspecting eqs.~(\ref{cumulant_ex_1}) and (\ref{cumulant_ex_2}), we find
\begin{eqnarray}
&&
-\frac{f\left< \Delta x(t)\right>}{k_BT} +\frac{f^2\left< \delta \Delta x(t)^2\right>}{2(k_BT)^2}
\label{cumulant_ex_3}
\\
&=&
\sum_{q=0}^N
\left[
-c_q\frac{F_q\left< \Delta X_q(t) \right>}{k_BT/N}
+
F_q^2\frac{\left< \delta \Delta X_q(t)^2 \right>}{2(k_BT/N)^2}
\right]
(h_{q,N}^\dagger)^2.
\nonumber 
\end{eqnarray}
If the FDR in the mode space, i.e., eq.~(\ref{FDR_mode_app}), holds, then we arrive at the FDR in real space $\left< \Lambda (t) \right>=0$.
It should be noted that eq.~(\ref{cumulant_ex_3}) is obtained under the assumption of a Gaussian distribution.

\subsection{Instantaneous response}
\label{cal_stochastic_respnse_integralbyparts}

Let us consider our motivation for the stochastic response function.

Looking back at eq.~(\ref{PDF_noise}), we have the path probability given $X_q(0)$, which is written with $X_q$ for $t\in [0,t]$:
\begin{eqnarray}
&&{\cal P}_q[X_q(\cdot)|X_q(0)]
\nonumber\\
&\sim&
J
\exp{}\Biggl(-\int_0^tdt'\, \frac{1}{4c_qk_BT\gamma_q(t')/N}  
\nonumber \\
&&
\times \left[ 
\gamma_q(t') \frac{dX_q(t')}{dt'}
+k_{q}(t') X_q(t') -F_q(t')
\right]^2 \Biggr),
\end{eqnarray}
where the Jacobian is denoted by $J=\left| \delta Z_q(t'')/\delta X_q(t') \right|$.
Note that we consider the overdamped Langevin eq.~(\ref{imag_EM_TDSA}), with which the Jacobian is specified as
\begin{eqnarray}
J=\exp{}\left[ \int_0^tdt'\, \frac{1}{2\gamma_q(t')} \frac{\partial^2{\cal E}_q[X_q,t']}{\partial X_q^2} \right].
\label{Jacob}
\end{eqnarray}
As well as for the case that ${\cal E}_q(X_q,t')$ is the harmonic potential, the Jacobian does not matter in the following calculations, because 
\begin{eqnarray}
\frac{\delta J}{\delta F_q(t)}
&=&
J\frac{1}{2\gamma_q(t)} \frac{\delta}{\delta F_q(t)}
\left[
\frac{\partial^2{\cal E}_q[X_q,t]}{\partial X_q^2} 
\right]
=0.
\label{Jacob_dev}
\end{eqnarray}
With eq.~(\ref{Jacob_dev}) in mind, we calculate
\begin{eqnarray}
&& \frac{\delta [ -\log{{\cal P}_q[X_q(\cdot)|X_q(0)]} ]}{\delta F_q(t)}
\nonumber \\
&=&
\frac{\delta}{\delta F_q(t)}
\Biggl(\int_0^tdt'\, \frac{1}{4c_qk_BT\gamma_q(t')/N}  
\nonumber \\
&&
\times \left[ 
\gamma_q(t') \frac{dX_q(t')}{dt'}
+\frac{\partial {\cal E}_q(X_q,t)}{\partial X_q} -F_q(t')
\right]^2 \Biggr)
\nonumber \\
&=&
-
\frac{N}{2c_qk_BT}  
\left[ 
\frac{dX_q(t)}{dt}
+\frac{1}{\gamma_q(t)}\frac{\partial {\cal E}_q(X_q,t)}{\partial X_q} -\frac{F_q(t)}{\gamma_q(t)}
\right].
\label{delSq_Fq}
\end{eqnarray}
Substituting eq.~(\ref{imag_EM_TDSA}) of motion into eq.~(\ref{delSq_Fq}) gives
\begin{eqnarray}
\frac{\delta [ -\log{{\cal P}_q[X_q(\cdot)|X_q(0)]} ]}{\delta F_q(t)}
&=&
-\frac{Z_q(t)}{2c_qk_BT\gamma_q(t)/N}.
\label{ProbXq_Qq}
\end{eqnarray}
In addition, noting ${\cal P}_q(X_q(t))=\int dX_q(0)\int {\cal D}X_q(s')\,{\cal P}_q[X_q(\cdot)|X_q(0)]$,
we discover
\begin{eqnarray}
\frac{\delta [ -\log{{\cal P}_q[X_q(\cdot)|X_q(0)]} ]}{\delta F_q(t)}
=
\frac{\delta [ -\log{{\cal P}_q(X_q(t))} ]}{\delta F_q(t)}
\label{PqXq_PqPath}
\end{eqnarray}
Combining eq.~(\ref{PqXq_PqPath}) with eqs.~(\ref{stochastic_response}) and (\ref{ProbXq_Qq}), we find that the right-hand side of eq.~(\ref{stochastic_HS_def}) is $Q(t)$ (eq.~(\ref{SP_Qq})).

Next, we consider the ensemble average relation $\left<R^*_q(t,t)\right>=R_q(t,t)$.
With the notation $\dot{X}_q(t)\equiv dX_q(t)/dt$, we transform the response function by integration by parts:
\begin{eqnarray}
&& \left<\frac{\delta  \dot{X}_q(t)}{\delta F_q(s)} \right>
\nonumber \\
&\equiv&
 \int dX_q(0)\,{\cal P}_q(X_q(0)) \int {\cal D}Z_q (t')\, \frac{\delta \dot{X}_q(t)}{\delta F_q(s)} {\cal P}_q[Z_q(\cdot)]
 \nonumber \\
&=&
 \int dX_q(0)\,{\cal P}_q(X_q(0)) \int {\cal D}Z_q(t')\, \frac{\delta \dot{X}_q(t)}{\delta Z_q(s)} {\cal P}_q[Z_q(\cdot)]
\nonumber \\
&=&
-\int dX_q(0)\,{\cal P}_q(X_q(0)) \int {\cal D}Z_q (t')\, \dot{X}_q(t)  \frac{\delta {\cal P}_q [Z_q(\cdot)]}{\delta Z_q(s)}
\nonumber \\
&=&
\left< \dot{X}_q(t)\frac{\delta}{\delta Z_q(s)} \left[ -\log{{\cal P}_q[Z_q(\cdot)]} \right] \right>,
\label{IntegralByPart}
\end{eqnarray}
where, on the third line of the right-hand side, the boundary terms that appear through integration by parts are assumed to be ignored.
The term in brackets on the last line resembles eq.~(\ref{stochastic_response}), but is still distinct.
Recalling eq.~(\ref{PDF_noise}), 
we apply functional differentiation to $-\log{{\cal P}_q[Z_q(\cdot)]}$ with respect to $Z_q(t)$.
Comparing the result with eq.~(\ref{ProbXq_Qq}), we find 
\begin{eqnarray}
\frac{\delta [ -\log{{\cal P}_q[X_q(\cdot)|X_q(0)]} ]}{\delta F_q(t)}
&=&
-
\frac{\delta [ -\log{{\cal P}_q[Z_q(\cdot)]} ]}{\delta Z_q(t)} 
\label{ProbXq_Zq_Fq}
\end{eqnarray}
with attention to the sign.
Application of eqs.~(\ref{stochastic_response}),\,(\ref{PqXq_PqPath}), and (\ref{ProbXq_Zq_Fq}) to eq.~(\ref{IntegralByPart}) verifies that $\left<R^*_q(t,t)\right>=R_q(t,t)$.

\end{document}